\documentclass[11pt,a4paper]{article}
\usepackage{jcappub}
\usepackage{bm}
\usepackage{amsmath}
\def\dd{\mathrm{d}}

\def\mcP{\mathcal{P}}
\def\mcR{\mathcal{R}}

\def\Mpl{M_{\rm Pl}}
\def\GeV{{\rm GeV}}

\newcommand{\para}[1]{\par\vspace{2mm}\noindent\textbf{{#1}}.---}

\definecolor{verde}{rgb}{0,0.5,0}

\usepackage{titlesec}

\title{Primordial Gravitational Waves\\ from Axion-Gauge Fields Dynamics}

\author{Emanuela Dimastrogiovanni$^{a}$, Matteo Fasiello$^{b}$ and Tomohiro Fujita$^{b}$}
\affiliation[a]{Department of Physics and School of Earth and Space Exploration,\\ Arizona State University, Tempe, AZ 85827, U.S.A.}
\affiliation[b]{Stanford Institute for Theoretical Physics and Department of Physics, Stanford University,\\ Stanford, CA 94306, U.S.A.}
\emailAdd{edimastr@asu.edu}
\emailAdd{matteorf@stanford.edu}
\emailAdd{tomofuji@stanford.edu}

\abstract{ Inspired by the chromo-natural inflation model of Adshead\&Wyman, we reshape its scalar content to relax the tension with current observational bounds. Besides an inflaton, the setup includes a spectator sector in which an axion and SU(2) gauge fields are coupled via a Chern-Simons-type term. The result is a viable theory endowed with an alternative production mechanism for gravitational waves during inflation. The gravitational wave signal sourced by the spectator fields can be much larger than the contribution from standard vacuum fluctuations, it is distinguishable from the latter on the basis of its chirality and, depending on the theory parameters values, also its tilt.  This production process breaks the well-known relation between the tensor-to-scalar ratio and the energy scale of inflation. As a result, even if the Hubble rate is itself too small for the vacuum to generate a tensor amplitude detectable by upcoming experiments, this model still supports observable gravitational waves.
}

\arxivnumber{1608.xxxx}

\begin{document}

\maketitle

\section{Introduction}
\label{sec:introduction}

\noindent Inflation \cite{inflation} is arguably the most convincing and observationally robust \cite{Ade:2015lrj} paradigm to date accounting for the origin of cosmic structures. While strongly supporting the picture of an exponential expansion in the very early universe, data have not yet been able to shed light on the microphysics of inflation. The importance of gaining insight into cosmic inflation cannot be overstated: inflation may have taken place at energy scales that are unlikely to ever be reached with particle colliders and is therefore to be considered a special portal for high energy physics.       \\
\indent All inflationary models support the production of a stochastic background of gravitational waves (GW) (see e.g. \cite{Guzzetti:2016mkm} for recent reviews). These arise from (tensorial) quantum fluctuations of space-time, whose wavelength are stretched by the inflationary expansion and freeze out on super-horizon scales\footnote{Alternative production mechanisms for primordial GW include, e.g., primordial magnetic fields \cite{Bonvin:2014xia}, phase transitions \cite{Krauss:1991qu}, topological defects \cite{Vachaspati:1984gt}. Gravitational waves are also produced in models that are alternative to inflation such as string gas cosmology \cite{Brandenberger:2006xi} and ekpyrotic scenarios \cite{Khoury:2001wf}. See also \cite{Dubovsky:2009xk} for GW in e.g. modified gravity setups.}. The search for primordial gravitational waves  has its primary focus on the polarization of the Cosmic Microwave Background (CMB)\footnote{A number of promising efforts for primordial GW detection are being directed, e.g., towards gravitational lensing effects of the tensor modes in the CMB \cite{lensingCMB}, in the galaxy distribution \cite{lss}, and in 21cm fluctuations \cite{21cm}, fossil effects in the CMB and LSS \cite{fossils}, direct searches through interferometers \cite{interfer}.}: it is well-known that GW leave an imprint in the CMB in the form of a specific (parity-odd) polarization pattern, the B-modes \cite{Bm}. Numerous earth-based and balloon-borne experiments are currently searching for B-modes and a new generation of experiments will become operational within the next decade \cite{newg}.    \\
\indent The amplitude of primordial B-modes is directly related to $r$, defined as the ratio between the inflationary tensor and scalar power spectra, $r\equiv \mathcal{P}_{h}/\mathcal{P}_{\zeta}$. The most recent bound, $r_{0.05\,\text{Mpc}^{-1}}<0.07$, is the result of the joint analysis of BICEP2/KECK and Planck data \cite{Array:2015xqh}. The sensitivity of upcoming CMB experiments (stage-IV) is expected to reach $\sigma(r)\sim 0.001$ \cite{Abazajian:2013vfg}. If GW are entirely sourced from vacuum fluctuations, $r$ directly quantifies the energy scale of inflation, $V_{\text{inf}}^{1/4}\approx 10^{16}\,\text{GeV}\left(r/0.01\right)^{1/4}$, as well as providing a measure of the inflaton field displacement during inflation, $\Delta\phi/M_{\text{Pl}}\gtrsim (r/0.01)^{1/2}$ (Lyth bound \cite{Lyth:1996im}), two crucial pieces of information for model building. Other standard predictions are a red-tilt for the tensor power spectrum, $n_{T}\simeq -2\epsilon_H=-r/8$, where $\epsilon_H\equiv-\dot{H}/H^2$, and a symmetry between the two tensor helicities (``even chirality''). A detection of primordial GW would then confirm or disprove these predictions. It is important to stress that, even if the amplitude of B-modes will turn out to be below the sensitivity threshold of upcoming experiments, these will provide bounds severely restricting the viable parameter space for inflationary models \cite{Lasky:2015lej}.\\

\indent A correct interpretation of B-mode measurements is clearly subject to a correct understanding of their source. In the context of inflation, generation by vacuum fluctuations is not the only known mechanism. Tensor perturbations may be produced from particle production, as proposed in a number of interesting scenarios \cite{pp}. They may also be sourced by scalar fluctuations of spectator fields, as it was shown in \cite{smallcs} for a scalar spectator with a small sound speed. In all of these cases, the one-to-one correspondence between the tensor-to-scalar ratio and the energy scale of inflation may not hold and the features predicted for the tensor power spectrum may differ from those of standard vacuum production. \\
\indent Another interesting scenario with an alternative GW generation mechanism (specifically from gauge fields - axion dynamics), is known as \textsl{chromo-natural inflation} (henceforth ``CNI'') \cite{allCNI} \footnote{See \cite{recentCNI} for more recent work on CNI. In particular, as we shall see, the model in \cite{recentCNI}e shares and indeed includes the field content we will explore in this paper, the main difference being the scales and parameter space domains at the center of our investigation. See also \cite{Bielefeld:2014nza} for more studies on the effects of gauge fields on primordial gravitational waves.}. CNI was initially motivated by the search for a solution to the inflationary $\eta$-problem, i.e. devising ways to protect the flatness of the inflationary potential from large quantum corrections. The model was inspired by \textsl{natural inflation} \cite{natural}, in which the expansion is driven by a pseudo-scalar field with a typical axion potential, $V(\phi)\sim 1+\cos(\phi/f)$, $f$ being the decay constant of $\phi$. The flatness of the potential is protected by the (nearly exact) axionic shift symmetry. Agreement with observations in natural inflation is achieved for $f\gtrsim M_{\text{Pl}}$ \cite{fmp}, which for various reasons is an undesirable constraint on the theory \cite{fmpt}. Notably, in CNI one is able to successfully inflate with a sub-Planckian axion decay constant \footnote{See \cite{more-proposal} for more proposal on how to realize axion inflation with $f\lesssim M_{\text{Pl}}$.}. What allows the background to inflate for a large enough number of e-folds in spite of a smaller $f$ (i.e. of a steeper potential), is the friction on the axion dynamics provided by a coupling with $SU(2)$ gauge fields, $\phi F_{\mu\nu}^{a}\tilde{F}^{\mu\nu,a}$. In the process, gauge fields fluctuations source tensor and scalar modes, strongly affecting the cosmological predictions for both sectors. Specifically, gauge field tensor modes experience a transient growth in one of their polarizations. This affects the freeze-out values of the corresponding helicities of the metric tensor, hence leading to production of chiral GW. Curvature fluctuations are also generated from the axion-gauge dynamics. Unfortunately, studies have shown that in CNI it is not possible to simultaneously satisfy the bounds on $r$ and on the scalar spectral index, $n_{S}$ \cite{allCNI}.      \\
\indent Being able to accommodate observational bounds both in tensor and scalar sectors while allowing auxiliary fields to play an important role in the production of cosmological perturbations is a common challenge for many of the aforementioned alternative GW production mechanisms. It is often the case that GW cannot be produced by the auxiliary sector to an observable level if not at the expense of generating too much scalar non-Gaussianity or predicting a spectral index of curvature fluctuations that is ruled out. A solution to this issue has often been to free the auxiliary sector of any direct coupling with the inflaton. This route is exemplified by the work of \cite{Namba:2015gja}.       \\
\indent In this work, we propose a set-up where the field content of CNI is confined to a spectator sector, i.e. distinct from the inflaton one. As we show, the model preserves the rich phenomenology of CNI while remaining observationally viable. The most interesting predictions is the production of chiral GW sourced by the non-Abelian gauge field to a level that, in an ample region of the parameter space, overcomes the amplitude of vacuum-generated GW. This while being in agreement with observations on curvature fluctuations and above detection thresholds for near-future B-modes experiments.  \\

\para{Outline}This paper is organized as follows. In \S\ref{sec:review} we review chromo-natural inflation.        In \S\ref{sec:our} we present our model, analyze the background and linear fluctuations evolution. We also  compute the power spectra of gravitational waves and curvature perturbations for a chosen set of the model parameters. In \S\ref{sec:discussion} we elaborate on our findings and perform a scan of the parameter space. We summarize our work and offer our conclusions in \S\ref{sec:conclusions}.

\section{Review of chromo-natural inflation}
\label{sec:review}

\noindent In CNI the coupling between the gauge fields and the pseudo-scalar inflaton allows the latter to inflate for a sufficient number of e-folds on a steeper potential, i.e. in the presence of a sub-Planckian decay constant ($f$). This can be viewed as an energy transfer from the axion to the gauge field sectors (specifically a loss of kinetic energy for the axion) or, equivalently, as a damping effect of the gauge field on the motion of the axion. In the following we briefly review the main features of the model along with the background and perturbation analysis, as performed in previous work \cite{allCNI}. The Lagrangian for CNI is given by 
\begin{equation}\label{cnii}
S_{\text{CNI}}=\int d^4x\sqrt{-g}\left[\frac{M_{\text{Pl}}^{2}}{2}R-\frac{1}{2}\left(\partial\chi\right)^{2}-U(\chi)-\frac{1}{4}F_{\mu\nu}^{a}F^{a\mu\nu}+\frac{\lambda\,\chi}{4f}F_{\mu\nu}^{a}\tilde{F}^{a\mu\nu}\right]\,,
\end{equation}
where $M_{\text{Pl}}$ is the reduced Planck mass, $R$ is the Ricci scalar, $\chi$ is a pseudo-scalar field (axion) with potential $U(\chi)$, $F_{\mu\nu}^{a}\equiv\partial_{\mu}A^{a}_{\nu}-\partial_{\nu}A^{a}_{\mu}-g \epsilon^{abc} A^{b}_{\mu}A_{\nu}^{c}$ is the field strength of an $SU(2)$ gauge field $A_{\mu}^{a}$ and $\tilde{F}^{a\mu\nu}\equiv\epsilon^{\mu\nu\rho\sigma}F^{a}_{\rho\sigma}/(2\sqrt{-g})$ its dual. We take the following ansatz for the vev of the gauge field\footnote{This isotropic configuration of the gauge field is known to be the attractor solution \cite{recentCNI}b.}
\begin{equation}
A_{0}^{a}=0\,,\quad\quad A_{i}^{a}=\delta_{i}^{a} a(t) Q(t)\,.
\end{equation}
From Einstein equations one finds \cite{allCNI}
\begin{eqnarray}
&&3M_{\text{Pl}}^{2}H^2=\frac{\dot{\chi}^2}{2}+U(\chi)+\frac{3}{2}\left(\dot{Q}+HQ\right)^{2}+\frac{3}{2}g^2 Q^4\,,\\\label{slow-roll1}
&&-2M_{\text{Pl}}^{2}\dot{H}=\dot{\chi}^2+\frac{2}{a^2}\left[\partial_{t}\left(aQ\right)\right]^2+2g^2Q^4\,.
\end{eqnarray}
Eq.~(\ref{slow-roll1}) can be rewritten in terms of the slow-roll parameters, $\epsilon_H\equiv -\dot{H}/H^2=\epsilon_{\chi}+\epsilon_{B}+\epsilon_{E}$ where $\epsilon_{\chi}\equiv\dot{\chi}^{2}/(2 H^2 M_{\text{Pl}}^{2})$, $\epsilon_{B}\equiv g^2 Q^4/(H M_{\text{Pl}})^{2}$ and $\epsilon_{E}\equiv (HQ+\dot{Q})^2/(HM_{\text{Pl}})^2$, all of which are assumed to be much smaller than unity. The equations of motion for axion and gauge field read
\begin{eqnarray}\label{axion-gauge-bck}
&& \ddot{\chi}+3H\dot{\chi}+U_{\chi}(\chi)=-\frac{3g\lambda}{f}Q^2 \left(\dot{Q}+HQ\right) \,,\\
&& \ddot{Q}+3H\dot{Q}+\left(\dot{H}+2H^2\right)Q+2g^2 Q^3=\frac{g\lambda}{f}\dot{\chi}Q^2    \,,
\label{gauge-bck}
\end{eqnarray}
where $U_{\chi}\equiv\partial_{\chi}U$. It is convenient to combine the parameters as follows: $\Lambda\equiv \lambda Q/f$ and $m_{Q}\equiv g\,Q/H$. In the slow-roll approximation ($\dot{H}\ll H^2$, $\ddot{\chi}\ll H\dot{\chi}$, $\ddot{Q}\ll H\dot{Q}$) and restricting the parameter space to $\Lambda\gg \sqrt{2}$ and $\Lambda\gg \sqrt{3}/m_{Q}$, the background gauge field that minimizes its effective potential is given by
\begin{equation}\label{sol-sr1}
Q_{\text{min}}=\left(\frac{-f\, U_{\chi}}{3g\lambda H}\right)^{1/3}\,.
\end{equation}
This also leads to
\begin{equation}\label{sol-sr2}
\xi\equiv \frac{\lambda}{2fH}\dot{\chi}\simeq m_{Q}+\frac{1}{m_{Q}}\,.
\end{equation}
\noindent We adopt the following decomposition for gauge field fluctuations (se e.g. \cite{allCNI}e) 
\begin{eqnarray} \label{dec1}
&&A_{0}^{a}=a (Y_{a}+\partial_{a}Y)\,,\nonumber\\
&&A_{i}^{a}=a\left[\left(Q+\delta Q\right)\delta_{ai}+\partial_{i}\left(M_{a}+\partial_{a}M\right)+\epsilon_{iac}\left(W_{c}+\partial_{c}W\right)+T_{ia}\right]\,,
\end{eqnarray}
where $Y_{i}$, $M_{i}$ and $W_{i}$ are transverse vector perturbations and $t_{ij}$ is a traceless and transverse tensor. The contributions of metric fluctuations to the linear equations of motion from scalar modes are subleading and can be safely neglected \cite{allCNI}e. Vectors perturbations have been previously shown to quickly decay on super-horizon scales \cite{allCNI} and they can therefore be safely neglected \footnote{The parameter space allows for a small region where the vector perturbations undergo a transient instability. This may lead to anisotropic signatures for the model. In the same region, however, scalar fluctuations may become unstable \cite{allCNI}.}. The non-zero components of the metric tensor are then given by
\begin{eqnarray}\label{dec2}
g_{00}=-a^2\,,\quad\quad g_{ij}=a^2\left(\delta_{ij}+h_{ij}\right)\,.
\end{eqnarray}
The $SU(2)$ gauge freedom allows to fix $W=W_{i}=0$. With this gauge choice and setting $k=k_z$, one can rewrite Eqs.~(\ref{dec1}-\ref{dec2}) as 
\begin{eqnarray}
&&A_{\mu}^{1}= a\left(0,Q+\delta Q+T_{+},T_{\times},0\right)    \,,\nonumber\\
&&A_{\mu}^{2}= a\left(0,t_{\times},Q+\delta Q-T_{+},0\right)     \,,\nonumber\\
&&A_{\mu}^{3}= a\left(\partial_{z}Y,0,0,Q+\delta Q+\partial_{z}^{2}M\right)     \,,\nonumber\\
&&g_{ij}=\begin{bmatrix}
    1+h_{+} & h_{\times} & 0  \\
      & 1-h_{+}  & 0 \\
      &  & 1
  \end{bmatrix}  \,.
\end{eqnarray}

\noindent The equations of motion for the tensor modes to leading order in slow-roll are   
\begin{eqnarray}\label{tensorf1}
&&\partial_x^2 \psi_{R,L}+\left(1-\frac{2}{x^2}\right)\psi_{R,L}=\frac{2\sqrt{\epsilon_{E}}}{x}\partial_x t_{R,L}+\frac{2\sqrt{\epsilon_{B}}}{x^2}\left(m_{Q}\mp x\right)t_{R,L}\,,\\
&&\partial_x^2 t_{R,L}+\left[1+\frac{2}{x^2}\left(m_{Q}\,\xi \mp x(m_{Q}+\xi)\right)\right]t_{R,L}=-\frac{2\sqrt{\epsilon_{E}}}{x} \partial_x \psi_{R,L}+\frac{2}{x^2}\left[(m_{Q}\mp x)\sqrt{\epsilon_{B}}+\sqrt{\epsilon_{E}}    \right]\psi_{R,L}\,,\nonumber
\end{eqnarray}
where derivatives are defined w.r.t. $x\equiv k/aH$. We have introduced the right and left helicities for the tensors, $\psi_{L,R}\equiv (a M_{\text{Pl}}/2)(h_{+}\pm i h_{\times})$ and $t_{L,R}\equiv a(T_{+}\pm i T_{\times})$.\\
\indent From Eqs.~(\ref{tensorf1}) one can verify that the tensor fluctuations of the gauge field freeze out at late times. These would instead decay on super-horizon scales if they were not coupled to the tensor modes of the metric. The effect of the gauge modes on the metric fluctuations is also important: it was shown \cite{allCNI}, as one may verify from (\ref{tensorf1}), that $t_{R}$ experiences a transient instability at around horizon crossing. This instability translates into an enhancement of the corresponding helicity of the tensor fluctuations of the metric, a larger freeze-out value for the latter and, therefore, a chiral GW signal. This is a distinctive signature of the model.    \\

\noindent The equations of motion for the scalar fluctuations have the following form (see e.g. \cite{allCNI}e) 
\begin{equation}\label{scalars1}
\partial_x^2 \Delta_{i}-2\, K_{ij} \partial_x \Delta_{j}+\left(\Omega^{2}_{ij}-\partial_x K_{ij}\right)\Delta_{j}=0\,,
\end{equation}
where the index $i$ runs from 1 to 3 and we have defined: $\delta\chi\equiv\Delta_{1}/a$, $\delta Q\equiv \Delta_{2}/(\sqrt{2}a)$, $ M\equiv (a g Q \Delta_{2}+\sqrt{k^2+2 a^2 g^2 Q^2}\Delta_{3})/(\sqrt{2} g a^2 k^2 Q)$. The non-zero entries of $K$ and $\Omega$ are   
\begin{eqnarray}\label{eee}
&&K_{12}=\frac{\Lambda \,m_{Q}}{\sqrt{2}x}     \,, \qquad  K_{13}=-    \frac{\,\Lambda\,m_{Q}^2}{\sqrt{2}x (x^2+2m_{Q}^{2})^{1/2}}   \,,        \nonumber \\
&&\Omega_{11}= 1 - \frac{2+\epsilon_\chi+\epsilon_B+\epsilon_E}{x^2} + \frac{U_{\chi\chi}}{H^2x^2}+ \frac{m_{Q}^{2}\,\Lambda^2}{x^2+2m_{Q}^{2}} \,,  \quad\quad\Omega_{12}=\frac{3\,\Lambda\,m_{Q}}{\sqrt{2}x^2}\left(1+\frac{2}{3}\epsilon_Q\right)\,, \nonumber \\
&&\Omega_{13}=  -\frac{\Lambda m_Q^2}{\sqrt{2}x^2 (x^2+2m_Q^2)^{1/2}}-\Lambda \frac{2x^4+3x^{2}m_Q^2+4m_Q^4}{\sqrt{2}x^2(x^2+2m_Q^2)^{3/2}}(1+\epsilon_Q),  \quad\Omega_{22}=  1+\frac{4m_{Q}^{2}-2m_Q\xi}{x^2}\,, \nonumber \\
&&\Omega_{33}=  1+\frac{4m_{Q}^{2} (x^{2}+m_{Q}^{2} )}{x^2 (x^{2}+2m_{Q}^{2} )} -\frac{2\xi m_{Q}}{x^{2}+2m_{Q}^{2}}+\frac{6 m_{Q}^{2}(1+\epsilon_Q)^2}{ (x^{2}+2m_{Q}^{2} )^2} \,,\nonumber \\
&&\Omega_{23}=  -2 \frac{m_{Q}-\xi}{x^2}\sqrt{x^{2}+2m_{Q}^2} \,,  
\end{eqnarray}
where we have introduced $\epsilon_Q\equiv \dot{Q}/QH$.
It was previously shown \cite{allCNI} that scalar fluctuations may suffer from a tachyonic instability on sub-horizon scales if $m_{Q}<\sqrt{2}$. It was also shown that the parameter space of the model does not allow to successfully accommodate the observational bounds on both scalar and tensor power spectra \cite{allCNI}. In the next section, we prove that this is no longer the case if one introduces an inflaton sector minimally coupled to the (spectator) chromo-natural sector, while retaining the rich phenomenological implications of the model.  


\section{The model}
\label{sec:our}

Introducing an additional scalar field ($\phi$) as the inflaton, one can decouple, up to gravitational interactions, the dynamics of $\phi$ from the axion$+$gauge field sector, from now on acting as spectators for inflation. As we will show, this scheme allows for a large portion of the parameter space where the effects of the spectator sector on the super-horizon curvature fluctuations are negligible. As a result, cosmological predictions are easily reconciled with observations while preserving the ability for the gauge field to source GW.  \\
\indent Let us consider for simplicity a single-field inflaton sector with a generic potential. The Lagrangian (\ref{cnii}) is then modified as
\begin{equation}
S=\int d^4x\sqrt{-g}\left[\frac{M_{\text{Pl}}^{2}}{2}R-\frac{1}{2}\left(\partial\phi\right)^{2}-V(\phi)-\frac{1}{2}\left(\partial\chi\right)^{2}-U(\chi)-\frac{1}{4}F_{\mu\nu}^{a}F^{a\mu\nu}+\frac{\lambda\,\chi}{4f}F_{\mu\nu}^{a}\tilde{F}^{a\mu\nu}\,.\right]
\end{equation}
We present below the evolution for the background and for the linear fluctuations. In our calculations, we assume a standard axion potential $U(\chi)=\mu^4\left[1+\cos(\chi/f)\right]$. For illustrative purposes, in \S\ref{sec:background}-\ref{sec:scalars} we adopt the following parameters:
\begin{align} \label{parameter set 1}
&g=1.11\times10^{-2},\quad \lambda=500,\quad  \chi_*=\frac{\pi}{2}f=6.28\times10^{16}\GeV,\\
&H_* = 1.28\times10^{13}\GeV,\quad \mu = 1.92\times10^{15}\GeV,
\notag
\end{align}
\noindent where asterisk $*$ denotes the value at the horizon crossing of the observed mode and $\mu$ is the overall amplitude of the axion potential. 
In \S\ref{sec:discussion}, predictions for two more examples, respectively with a lower energy scale and a smaller coupling constant $\lambda$, are presented.

\subsection{Background evolution}
\label{sec:background}

\noindent Let us explore the background dynamics of our model. The $00$ component of Einstein equation is given by
\begin{eqnarray}
&&3M_{\text{Pl}}^{2}H^2=\frac{\dot{\phi}^2}{2}+V(\phi)+\frac{\dot{\chi}^2}{2}+U(\chi)+\frac{3}{2}\left(\dot{Q}+HQ\right)^{2}+\frac{3}{2}g^2 Q^4\,.
\label{our Friedmann}
\end{eqnarray}
The momentum constraint reads $\epsilon_H \equiv -\dot{H}/H^2=\epsilon_{\phi}+\epsilon_{\chi}+\epsilon_{B}+\epsilon_{E}$,
with $\epsilon_{\phi}\equiv\dot{\phi}^{2}/(2H^2 M_{\text{Pl}}^{2})$. As one would expect, the background evolution of $\chi$ and $Q$ is described by the same equations as in CNI, (\ref{axion-gauge-bck}) and \eqref{gauge-bck}. \\
\indent For $\Lambda\gg \sqrt{2}$ and $\Lambda\gg \sqrt{3}/m_{Q}$, the slow-roll solutions in (\ref{sol-sr1}) and (\ref{sol-sr2}) apply for the gauge field and the axion. We set up the initial conditions for the background evolution in a regime where the above conditions on $\Lambda$ are satisfied\footnote{We stress here that in CNI these initial conditions are a necessary requirement for the axion to be able to drive a sufficiently long phase of slow-roll inflation. While, for the sake of simplicity, we adopt the same set of initial conditions in our model, we are not strictly bound to this choice and it would be interesting to also investigate different ones.}. We also assume that the inflaton potential dominates the total energy density of the universe and its quantum fluctuations are the most important contributions to the curvature perturbations in the CMB observational window (we will return to the latter point in Sec.~\ref{sec:scalars}, where we check the consistency of this assumption within our parameter space). In this case, one can keep track of the evolution of $\epsilon_\phi$ by assuming $\eta_\phi\equiv \Mpl^2V''(\phi)/V(\phi)$ is constant and $\epsilon_{\phi}\simeq \epsilon_{H}$. Under these premises, one can show that
$\dot{\epsilon}_\phi=(4\epsilon_\phi-2\eta_\phi)H\epsilon_\phi$ and that the scalar spectral index $n_s$ is given by $n_s-1=-6\epsilon_\phi+2\eta_\phi$. $\epsilon_\phi$ then satisfies
\begin{equation}
\dot{\epsilon}_\phi =\left(4\epsilon_\phi-6\epsilon_{\phi*}+1-n_{s*}\right)H\epsilon_\phi,
\qquad (\eta_\phi=\text{const.})
\label{EoM for epphi}
\end{equation}
where the central value for the spectral index reported by the Planck mission is $1-n_{s*}\approx 0.032$ \cite{Ade:2015lrj}.
In this way, we can keep the analysis as general as possible letting the inflaton sector unspecified. It is straightforward to generalize our study to account for specific models.

%
\begin{figure}[tbp]
    \hspace{-2mm}
  \includegraphics[width=70mm]{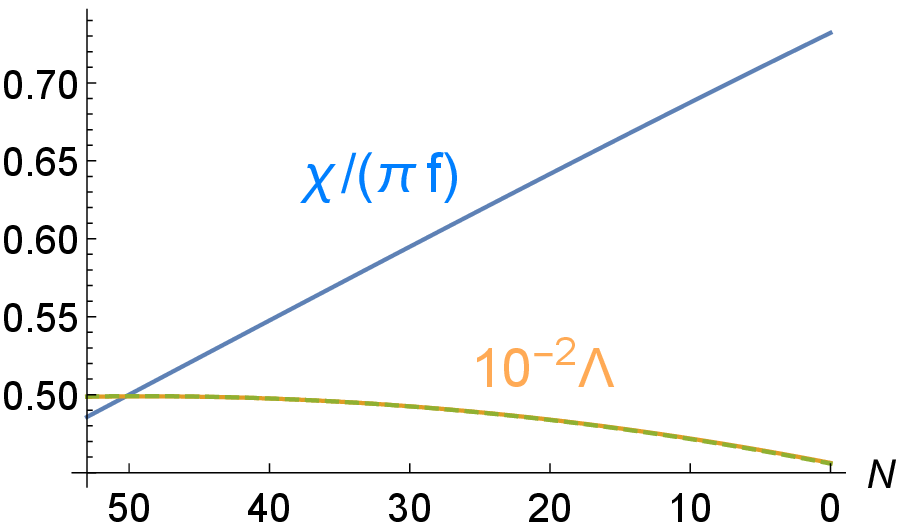}
  \hspace{5mm}
  \includegraphics[width=70mm]{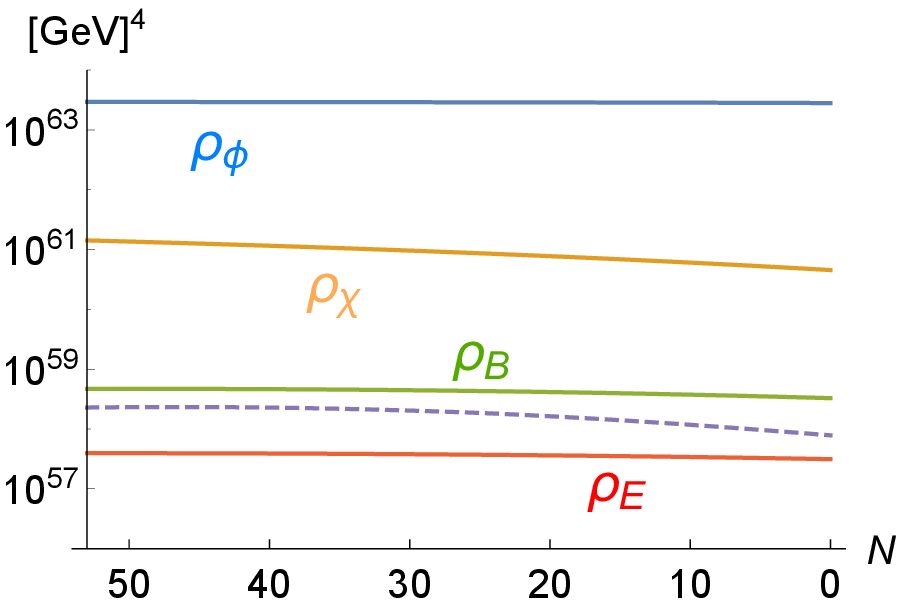}
  \caption
 {{\bf(Left panel)} The background evolution of the axion $\chi(t)/f$ (blue) and the gauge field $10^{-2}\Lambda=5Q(t)/f$ (yellow) are shown. The slow-roll expression $Q_{\min}(\chi)$, into which the numerically obtained $\chi(t)$ and $H(t)$ are plugged, is also plotted (green dashed line) and it coincides with the numerical one. 
{\bf(Right panel)} The energy density of the inflaton $\rho_\phi$ (blue),
the axion $\rho_\chi$ (yellow), the gauge field $\rho_B\equiv 3g^{2}Q^4/2$ (green), $\rho_E \equiv 3(\dot{Q}+HQ)^2/2$ (red) and the tensor fluctuation $\rho_{t_R}$ defined in eq.~\eqref{rho tR} (purple dashed) are shown. The inflaton always dominates the total energy density of the universe during inflation.}
 \label{chi and rho}
\end{figure}
%
%
\begin{figure}[tbp]
    \hspace{-2mm}
  \includegraphics[width=70mm]{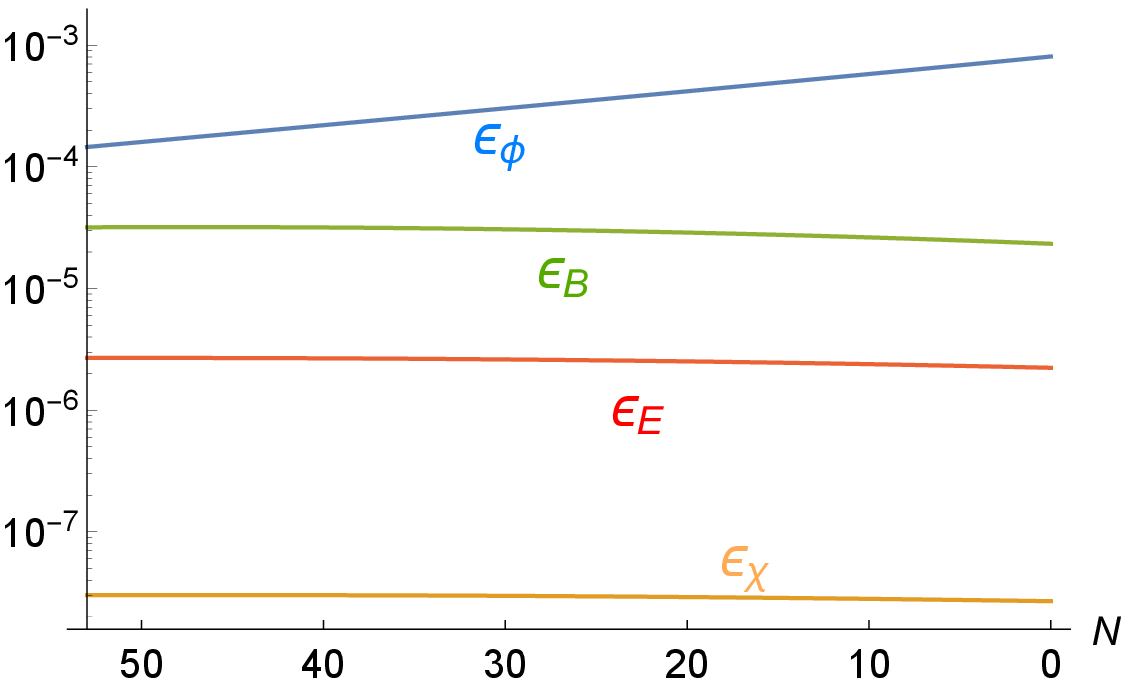}
  \hspace{5mm}
  \includegraphics[width=70mm]{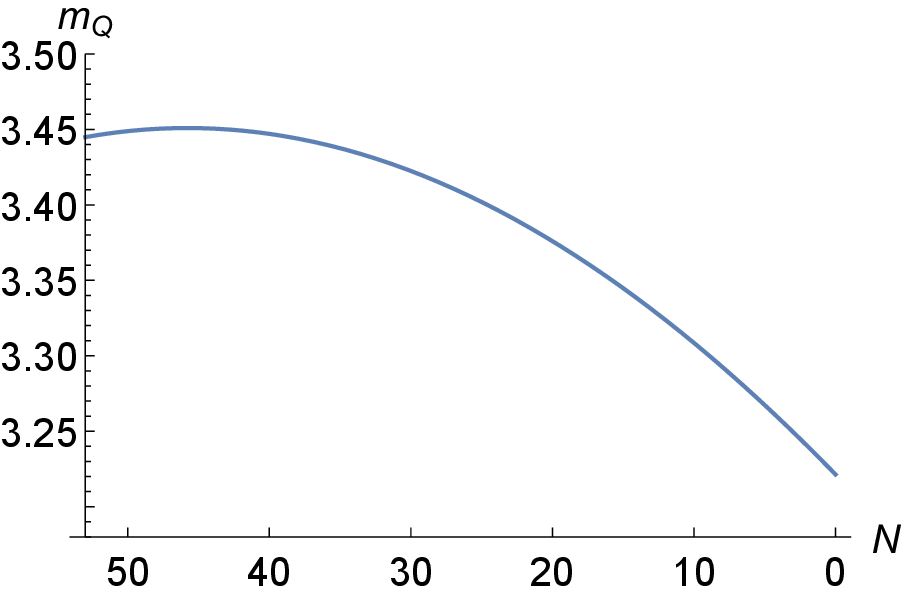}
  \caption
 {{\bf(Left panel)} The contributions to $\epsilon_H$ from the various components are shown, $\epsilon_H =\epsilon_{\phi}+\epsilon_{\chi}+\epsilon_{B}+\epsilon_{E}$. $\epsilon_\phi$ dominates $\epsilon_H$ and determines the evolution of $H$.
{\bf(Right panel)} $m_Q\equiv g Q/H$ is plotted. The shape of the resultant power spectrum of GW is basically determined by the time evolution of $m_Q$ as shown in \S\ref{sec:tensors}.}
 \label{epsilon and mQ}
\end{figure}
%
In Fig.~\ref{chi and rho}, we plot the evolution of the axion $\chi$, gauge field $Q$ and the energy density of the each component.
In the left panel, one sees that $\chi$ monotonically increases by slowly rolling down its potential and the evolution of Q is well described by its
slow-roll expression \eqref{sol-sr1}. The right panel indicates
that the energy densities of the axion $\rho_\chi$ and the gauge field $\rho_Q=\rho_B+\rho_E$ are much smaller than that of the inflaton, as it should be expected for a spectator sector. 
 Note that we consider
a CMB mode, $k_{*}$, exiting the horizon at $N_*=50$ for simplicity, where  $N(t)$ is the number of e-folds between a given time and the end of inflation.
With the parameters in \eqref{parameter set 1}, $\chi_*/f=\pi/2$ and $\Lambda_*=50$ at $N=N_{*}$.

In Fig.~\ref{epsilon and mQ}, we plot $\epsilon$ and $m_Q$ parameters. 
In the left panel, one can see that $\epsilon_\phi$ remains much larger than the others, consistently with the above
treatment of $\epsilon_\phi$ in Eq.~(\ref{EoM for epphi}). As shown in the right panel, $m_Q$ slowly decreases while remaining large enough to avoid any scalar instability. Since $\chi$ is at the inflection point of its potential ($U_{\chi\chi}=0$) at $N=50$ for the current parameter set, $m_Q\simeq g Q_{\min}/H \propto U_\chi^{1/3}/H^{4/3}$ reaches its peak value slightly after that due to the decreasing $H$.
As we will see in the next subsection, the GW power spectrum directly reflects this behavior of $m_Q$. 

\subsection{Tensor fluctuations}
\label{sec:tensors}

In this subsection, we calculate the tensor fluctuation both analytically and numerically.
To leading order in slow-roll, the equations of motion for the tensor modes are identical to (\ref{tensorf1}).

\begin{figure}[tbp]
    \hspace{-2mm}
  \includegraphics[width=70mm]{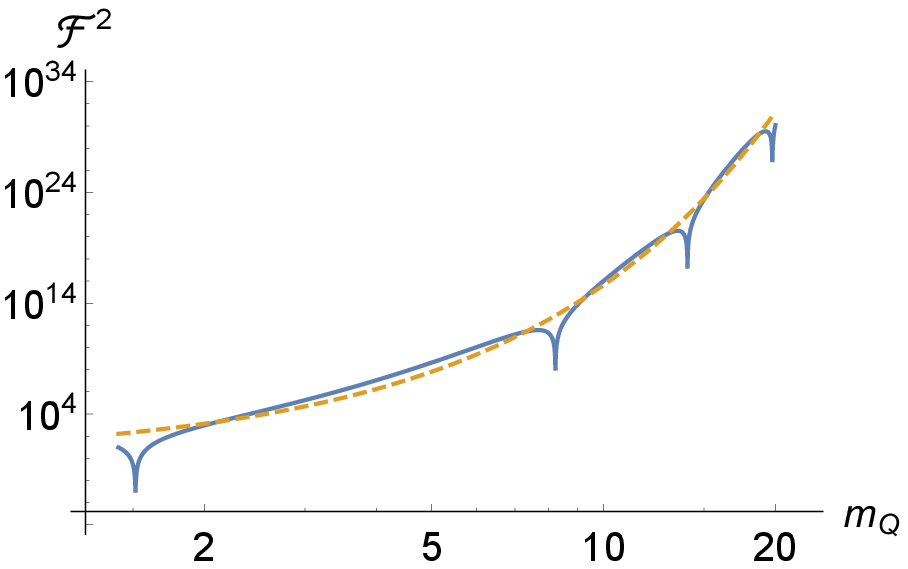}
  \hspace{5mm}
  \includegraphics[width=70mm]{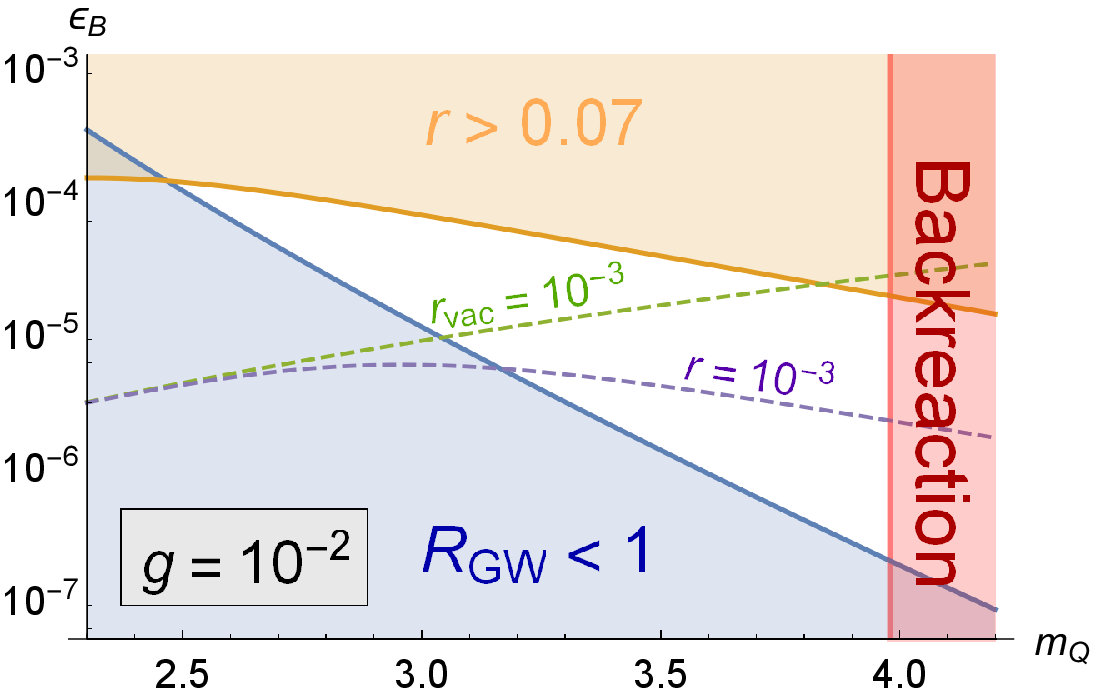}
  \caption
 {{\bf(Left panel)} $\mathcal{F}^2(m_Q)$ defined below \eqref{PQh} is evaluated under the slow-roll approximation. It is  roughly approximated by $\exp(3.6m_Q)$ (yellow dashed line). 
{\bf(Right panel)} The constraint on $m_Q$ and $\epsilon_B$ for $g=10^{-2}$ based on the analytic estimation of the sourced GW. In the blue shaded region, the sourced GW is smaller than one from the vacuum fluctuation ($R_{\rm GW}<1$). In the upper shaded region, the observational upper bound on the tensor-to-scalar ratio ($r<0.07$) is violated. The green and purple dashed lines denote $r_{\rm vac}=10^{-3}$ and $r=10^{-3}$, respectively. In the red shaded region, the backreaction on the EoM for $Q$ and $\chi$ is significant as discussed in \S\ref{sec:backreaction}.}
 \label{F and epB}
\end{figure}

First, we analytically estimate the amplitude of gravitational waves by using the Green's function method. Since only the right-helicity tensor mode of the gauge field $t_R$ is amplified by the instability and it sources only $\psi_R$, we focus on these modes. Assuming $m_Q$ and $\xi$ are constant,
one finds the homogeneous solution for $t_R$ is given by~\cite{recentCNI}e
\begin{equation}
t_R(x) = \frac{1}{\sqrt{2k}} i^{\beta}\,W_{\beta,\alpha}(-2ix),
\label{analytic tR}
\end{equation}
where $W_{\kappa,\mu}(z)$ is the Whittaker function, $\alpha \equiv -i\sqrt{2m_Q \xi -1/4}$ and $\beta \equiv -i(m_Q+\xi)$.
Here we have used the WKB solution in the sub-horizon limit, $t_R(x\to \infty)=(2k)^{-1/2} (2x)^\beta e^{i x}$, as the initial condition. Integrating the Green's function of $\psi_R$ multiplied by the source term with this $t_R$, one obtains the inhomogeneous solution for $\psi_R$ in the super-horizon limit as
\begin{equation}
\lim_{x\to0} \psi_R^{(s)}(x)
=\frac{1}{\sqrt{2k} x}\Big[\mathcal{F}_E \sqrt{\epsilon_E}+\mathcal{F}_B \sqrt{\epsilon_B}\Big],
\end{equation}
where $(s)$ denotes the ``sourced" solution and we have defined 
\begin{align}
\mathcal{F}_E\equiv
& \frac{\pi i^{\beta+1}\cos^{-1}(\pi\alpha)[16\alpha^4-40\alpha^2+9]^{-1}}
{\Gamma(1-\beta)\Gamma(\frac{1}{2}-\alpha-\beta)\Gamma(\frac{1}{2}+\alpha-\beta)}
\Bigg[16 (4\alpha^2-8\beta-1)\Gamma^2(1-\beta)
\notag\\
&-\left[16\alpha^4+8\alpha^2(8\beta-5)+16\beta(8\beta-1)+9\right]
\Gamma\left(\frac{1}{2}-\alpha-\beta\right)\Gamma\left(\frac{1}{2}+\alpha-\beta\right)
\notag\\
&+(16\alpha^4-40\alpha^2+9)\Gamma(1-\beta)\Gamma(-\beta)\Bigg],
\\
\mathcal{F}_B \equiv 
&\frac{\pi i^{\beta}\cos^{-1}(\pi\alpha)[16\alpha^4-40\alpha^2+9]^{-1}}
{\Gamma(1-\beta)\Gamma(\frac{1}{2}-\alpha-\beta)\Gamma(\frac{1}{2}+\alpha-\beta)}
\Bigg[8i \left[4(m_Q+i)\alpha^2-8m_Q\beta-9i-m_Q\right]\Gamma^2(1-\beta)
\notag\\
&-(4\alpha^2+8\beta-1)(4\alpha^2+8im_Q\beta-9)
\Gamma\left(\frac{1}{2}-\alpha-\beta\right)\Gamma\left(\frac{1}{2}+\alpha-\beta\right)
\notag\\
&+i(16\alpha^4-40\alpha^2+9)\Gamma(1-\beta)\Gamma(-\beta)\Bigg].
\end{align}
With these solutions, the power spectrum of the sourced GW in the super-horizon limit reads
\begin{align} \label{PQh}
\mcP_h^{(s)}(k) &= \frac{H^2}{\pi^2 \Mpl^2}\left|\sqrt{2k}x \lim_{x\to0}\psi_R^{(s)}(x)\right|^2
=\frac{\epsilon_B H^2}{\pi^2 \Mpl^2}\mathcal{F}^2,
\end{align}
with $\mathcal{F}^2 \equiv \left|\mathcal{F}_B+\sqrt{\epsilon_E/\epsilon_B}\mathcal{F}_E\right|^2$.
With the slow-roll equations $\xi\simeq m_Q+m_Q^{-1}$ and $\sqrt{\epsilon_E/\epsilon_B}\simeq m_Q^{-1}$, $\mathcal{F}^2$ can be written as a function of $m_Q$ and is plotted in Fig.~(\ref{F and epB}). Several dips seen in the plot are caused by cancellations between $\mathcal{F}_B$ and $\sqrt{\epsilon_E/\epsilon_B}\mathcal{F}_E$.%
\footnote{These dips can also be seen in the numerical solution with the time-dependent background quantities.} 
%
%

The ratio between the power spectrum of the sourced GW and that from the vacuum fluctuation is given by
\begin{equation}
R_{\rm GW} \equiv \frac{\mcP_h^{(s)}}{\mcP_h^{\rm vac}}
=\frac{\epsilon_B}{2}\mathcal{F}^2.
\end{equation}
The tensor-to-scalar ratio $r$ in our model is analytically estimated as
\begin{equation}
r=\frac{\mcP_h^{\rm vac}+\mcP_h^{\rm Q}}{\mcP_\zeta}
=\frac{2g^2\epsilon_B}{\pi^2 m^4_Q \mcP_\zeta}\left( 1+R_{\rm GW}\right),
\end{equation}
where $\mcP_\zeta$ is the power spectrum of the curvature perturbation and we have used $H^2/\Mpl^2 = g^2\epsilon_B/m^4_Q$.
We are interested in the parameter region where the sourced GW is larger than the vacuum GW, namely $R_{\rm GW}>1$, and $r$ satisfies the observational upper bound, $r\le 0.07$. This parameter space is shown in the right panel of Fig.~(\ref{F and epB}).
The white region in the plot is further separated into the following three.
(i) Above the green dashed line: both the sourced GW and vacuum GW can be
detected with near-future observations. (ii) Between the green and purple dashed lines: only the sourced GW can be detected while the vacuum GW are too weak to be observed. (iii) A sensitivity higher than $\sigma_{r}\sim10^{-3}$ is required to detect the signal.

It should be noted that the above estimates assume that $m_Q$ and $\xi\simeq m_Q+m_Q^{-1}$ are constant and employs the analytic solution in Eq.~\eqref{analytic tR}. Since the time evolution of $m_Q$ and $\xi$ can be significant depending on the parameters and Eq.~\eqref{analytic tR} may deviate from the numerical one by $\mathcal{O}(1)$ factors, numerical calculations are generally needed to precisely evaluate $\mcP_h$. However, the above analytic estimation and the right panel of fig.~\ref{F and epB} are very useful for a cross-check with numerical results and for parameters choice. \\

\begin{figure}[tbp]
    \hspace{-2mm}
  \includegraphics[width=70mm]{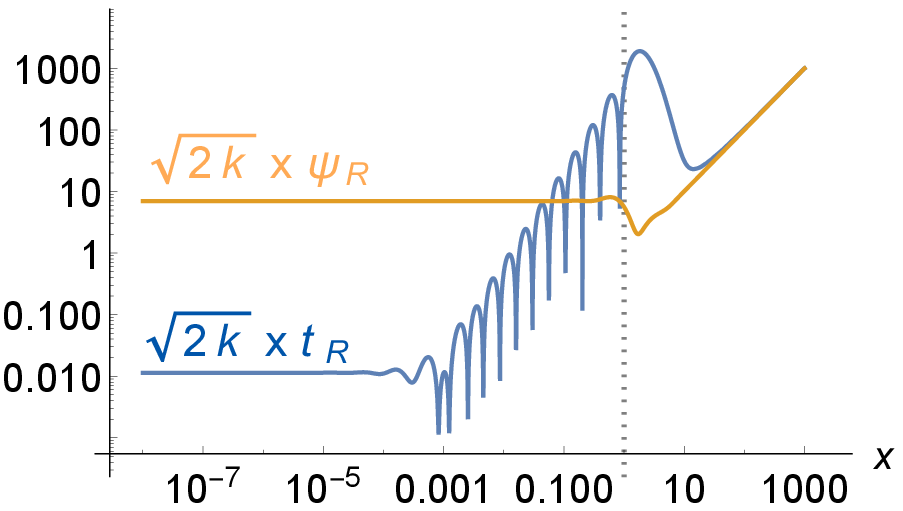}
  \hspace{5mm}
  \includegraphics[width=70mm]{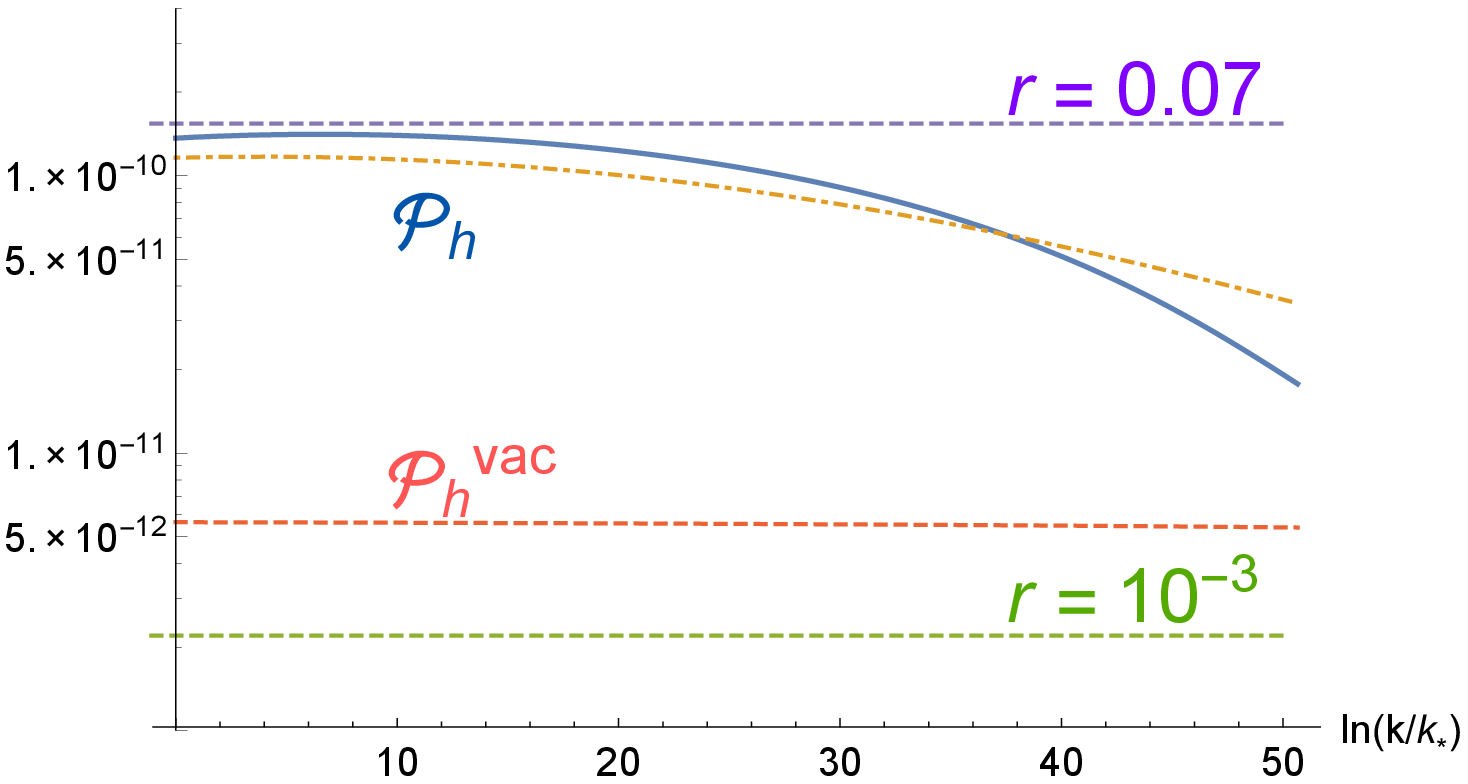}
  \caption
 {{\bf(Left panel)} The mode functions of the tensor fluctuation with the right helicity, $\psi_R(x)$
and $t_R(x)$, normalized by $\sqrt{2k}x$ for $k=k_*$. $t_R$ reaches its peak
and sources $\psi_R$ slightly before horizon crossing ($x=1$; vertical dotted line). After horizon crossing, $\psi_R$ remains constant.
{\bf(Right panel)} The resultant $\mcP_h$ (blue), $\mcP_h$ contributed only by the vacuum fluctuation
(red) and $\mcP_h$ estimated by the analytic expression (yellow dashed) are shown. The green and purple lines denote $r=10^{-3}$ and $0.07$, respectively. $\mcP_h$ is substantially enhanced by the axion-gauge spectator sector. }
 \label{GW fig1}
\end{figure}

Next, we numerically solve the equations for the tensor fluctuation (\ref{tensorf1})
with the time-dependent background quantities, $m_Q(t)$ and $\xi(t)$, to derive the power spectrum of GW. The results are shown in Fig.~\ref{GW fig1}. In the left panel, we plot the time evolution of  $\psi_R$ and $t_R$ for $k=k_*$. 
One can see that $t_R$ is amplified by the instability and reaches its peak value slightly before horizon crossing. Being sourced by $t_R$, $\psi_R$
is also amplified at the same time. After horizon crossing, the source effect from $t_R$ weakens and $\psi_R$ becomes constant. Note that the mode functions are normalized such that massless vacuum fluctuations are unity outside the horizon and thus the enhancement of the GW, $\sqrt{2k} x \psi_R >1$, is visible. Although $t_R$ starts decaying on super-horizon scales, it is in turn sourced by $\psi_R$ and becomes constant, with a much smaller amplitude. $t_R$ eventually decays when the background $Q$ decays and $t_R$ and $\psi_R$ decouple.
In the right panel of Fig.~(\ref{GW fig1}), we show the GW power spectra.
One can see that the resultant $\mcP_h$ is considerably larger than $\mcP_h^{\rm vac}$, while not exceeding the observational upper bound ($r=0.07$) for $k\sim k_*$. \\
\indent Plugging the numerically obtained $m_Q(t)$ and $\epsilon_B(t)$ at the time of horizon-crossing of each mode into \eqref{PQh}, one obtains the yellow dashed line in the right panel of Fig.~\ref{GW fig1}. This shows a reasonable agreement with the full numerical result and validates the parameter choice based on the analytic expression (see the right panel of Fig.~\ref{F and epB}).

\subsection{Scalar fluctuations}
\label{sec:scalars}

In this subsection, we consider the scalar fluctuations in the axion-gauge sector $\Delta_i$ ($i=1,2,3$) and their effect on the curvature perturbation $\zeta$.
First we numerically compute $\Delta_i$ by neglecting the metric fluctuations: although these couple $\delta\phi$ and $\Delta_i$, their effect is suppressed by the slow-roll parameters and can be ignored in the computation of $\Delta_i$. Next we evaluate the inflaton perturbation $\delta\phi^{(s)}$ which is induced by $\Delta_i$ through the gravitational coupling and confirm that it is negligible compared to the intrinsic one $\delta\phi^{(\rm vac)}$ from the vacuum fluctuation. Finally, we estimate the curvature perturbation $\zeta_\chi$ which is directly produced by the density perturbation of the axion, $\delta\rho_\chi$, at the end of inflation. \\

\begin{figure}[tbp]
    \hspace{-2mm}
  \includegraphics[width=70mm]{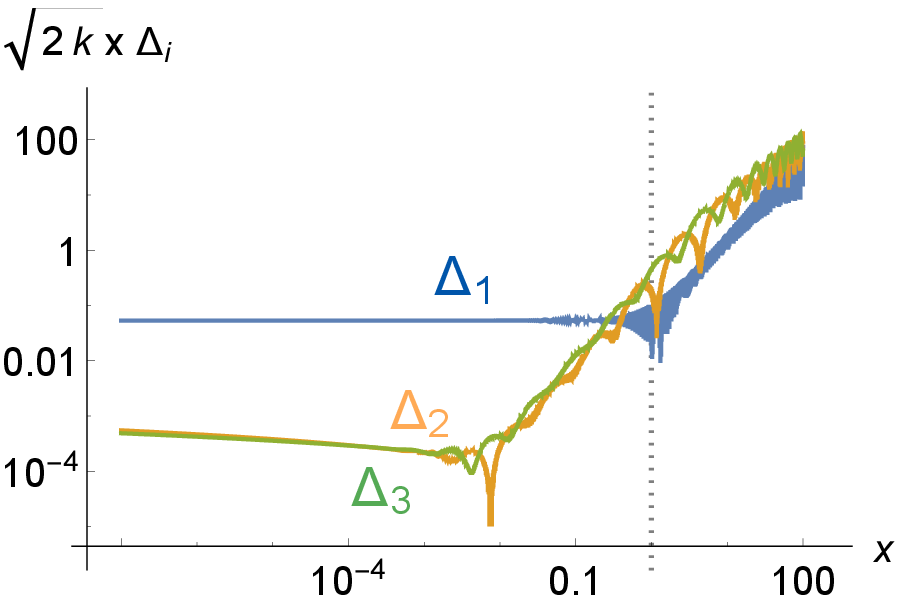}
  \hspace{5mm}
  \includegraphics[width=70mm]{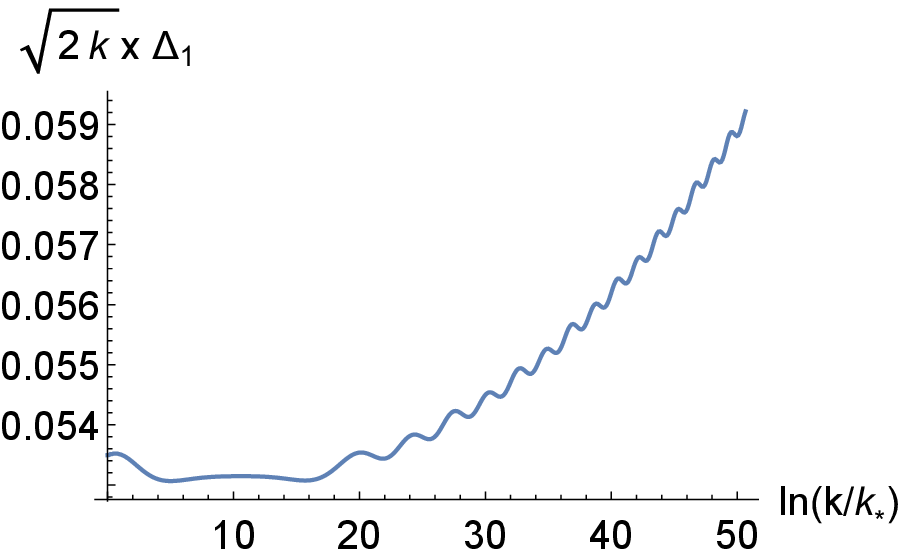}
  \caption
 {{\bf(Left panel)} The numerically obtained scalar perturbations in the axion-gauge sector, $\Delta_i\ (i=1,2,3)$. The axion fluctuation $\Delta_1$
freezes after horizon crossing, while its amplitude is much smaller than
in the standard massless case, $\sqrt{2k}x \Delta=1$. The scalar modes of the gauge field, $\Delta_2$ and $\Delta_3$, evolve on super-horizon scales depending on the axion mass $U_{\chi\chi}$, but their amplitudes are negligible.
{\bf(Right panel)} The spectrum of $x\Delta_1$ in the super-horizon limit. The amplitude is $\mathcal{O}(10^{-1})$ and $\delta\chi$ induces only negligible $\delta\phi$.}
 \label{Delta_and_Ps}
\end{figure}

The linearized equations of motion for the scalar fluctuation in the axion-gauge sector $\Delta_i$  are the same as in Eqs.(\ref{scalars1}), with the matrices $K$ and $\Omega$ equal to those in  Eqs.~(\ref{eee}), except for an $\epsilon_\phi$ added in the numerator of the second term in $\Omega_{11}$.
With the initial condition $\sqrt{2k}\Delta_j=1, \sqrt{2k}\partial_x\Delta_j=i, (j=1,2,3)$ at $x=2\times 10^4$ \cite{allCNI}e, we numerically compute $\Delta_i$ and plot them in Fig.~\ref{Delta_and_Ps}.
On super-horizon scales, the axion perturbation $\Delta_1\equiv a\delta\chi$ freezes out with a smaller amplitude than in the standard massless case ($\sqrt{2k}x \Delta=1$). The amplitudes of the gauge field perturbations $\Delta_2$ and $\Delta_3$ are even smaller and they eventually decay as $\Delta_{2,3}\propto a^{-1}$ after the slow-roll of $\chi$ and $Q$ terminates. Note that, in the current case, $m_Q$ is sufficiently large, $m_Q\simeq 3 >\sqrt{2}$, and no scalar instability takes place (see fig.~\ref{epsilon and mQ}).\\
%
%

Now let us consider the $\Delta_i$ contributions to the curvature perturbation $\zeta$ through the inflaton perturbation $\delta\phi^{(s)}$.  Although 
the inflaton and the axion-gauge sector are decoupled at the background level,  perturbations are coupled due to the metric fluctuation.
Since $\Delta_1 \gg \Delta_2, \Delta_3$, we focus on the contribution of $\Delta_1$.
Taking into account the gravitational coupling, one finds the equation of motion for $\delta\phi$ is given by (see e.g. \cite{Namba:2015gja})
\begin{equation}
\left(\partial_x^2+1 -\frac{2}{x^2}\right) \Delta_\phi 
= \frac{6\sqrt{\epsilon_\phi \epsilon_\chi}}{x^2} \Delta_1+
\mathcal{O}(\Delta_2,\Delta_3),
\end{equation}
where $\Delta_\phi\equiv a\delta\phi$ and the inflaton is approximated to be massless. Although we can numerically solve this equation with great accuracy, an analytic estimation under the approximation that $\epsilon_\phi,
\epsilon_\chi, x\Delta_\chi\simeq const.$ suffices for our purpose.
 With the Green's function method, one finds the inhomogeneous
solution for $\Delta_\phi$ in the super-horizon limit is given by
\begin{equation}
\lim_{x\to0}\sqrt{2k} x \Delta_\phi^{(s)} \simeq -2\sqrt{\epsilon_\phi \epsilon_\chi}
\ln (x) \left( \sqrt{2k}x\Delta_\chi \right),
\end{equation}
where $(\sqrt{2k}x\Delta_\chi)$ is treated as a constant. One should evaluate the r.h.s at the end of inflation, if $x\Delta_\chi$ remains constant until then. If not, the r.h.s. should be evaluated when $x \Delta_\chi$ starts decaying during inflation. With the chosen parameter set (\ref{parameter set 1}), we have $\epsilon_\phi =\mathcal{O}(10^{-4}), \epsilon_\chi= \mathcal{O}(10^{-8}), \sqrt{2k}x\Delta_\chi = \mathcal{O}(10^{-1})$ (see Fig.~\ref{Delta_and_Ps}) and $-\ln(x)\simeq 50$ for $k\sim k_*$. 
Therefore $\delta\phi^{(s)}/\delta\phi^{(\rm vac)}=\mathcal{O}(10^{-5})$
and the contribution to the curvature perturbation from the axion-gauge sector
through $\delta\phi^{(s)}$ is completely negligible\footnote{Since $\delta\phi^{(\rm vac)}$ and $\delta\phi^{(s)}$ are uncorrelated, the contribution from the latter to the curvature power spectrum $\mcP_{\zeta}$ is $\mathcal{O}(10^{-10})$ times smaller than the former one.}.\\

The axion perturbation $\delta\chi$ and the gauge field perturbation $\delta Q, M$ lead to  density fluctuation $\delta\rho_\chi$ and $\delta\rho_Q$ and can therefore also directly contribute to the curvature perturbation. However, the magnitude of these contributions may largely depend on the evolution of the axion-gauge sector after inflation. For instance, if $\chi$ dominates the universe after the inflaton decays, $\delta\chi$ can significantly contribute to the curvature perturbation similarly to what happens in the well-known \textsl{curvaton mechanism} \cite{curvatonm}. In this paper, for simplicity, we consider scenarios in which the energy fractions of $\chi$ and $Q$ decrease during or after inflation and the dominant contribution to $\zeta$ is due to $\delta\phi$. It is beyond the scope of this paper to fully investigate such a possibility by specifying the evolution of $\phi$ and $\chi$ after inflation. However, let us quickly assess the size of $\zeta$ induced by $\delta\rho_\chi$ at the end of inflation to get an insight. 
It is convenient to define the ratio between the contributions of $\delta\phi$ and $\delta\chi$ to $\zeta$ during inflation as
\begin{equation}
\mathcal{A}\equiv\left|\frac{\zeta_\chi}{\zeta_\phi}\right| =\left|\frac{\delta\rho_\chi}{\delta\rho_\phi}\right|
\simeq \frac{U\chi \delta\chi}{V_\phi \delta\phi}
\simeq \frac{U_\chi}{3\sqrt{2\epsilon} \Mpl H^2} \left( \sqrt{2k} x \Delta_1\right)
\,,
\end{equation}
where we have neglected the kinetic energy of $\phi$ and $\chi$, used the slow-roll approximation $\delta\rho_\phi \simeq 3\sqrt{2\epsilon} \Mpl H^2 \delta\phi$ and assumed the standard massless vacuum fluctuation $\sqrt{2k} x \Delta_\phi=1$ for $\delta\phi$ on super-horizon scales.
For $k\simeq k_*$, we find $\mathcal{A} \simeq 0.3$ at $N=0$ in the current case. Therefore the axion perturbation may produce (less/more than) $10\%$ of $\mcP_\zeta$ if the energy fraction of $\chi$ is unchanged (decreases/increases) after inflation.
Regarding the spectral index $n_s-1$, differentiating the log of $\mcP_\zeta= \mcP_\zeta^{(\phi)}(1+\mathcal{A}^2)$ w.r.t. $\ln k$, one finds that the additional contribution to $n_s -1$ is given by $(1+\mathcal{A}^2)^{-1}\dd \mathcal{A}^2/\dd\ln k$ and it is $\mathcal{O}(10^{-4})$ in the current case.
Although in this case the spectrum of $\delta\chi$ is nearly flat for scales in the vicinity $k=k_{*}$ because of our choice of the initial condition $\chi_*/f=\pi/2$, it can be much larger, and either negative or positive depending on the parameters values.


\subsection{Backreaction and consistency}
\label{sec:backreaction}

%
\begin{figure}[tbp]
    \hspace{-2mm}
  \includegraphics[width=70mm]{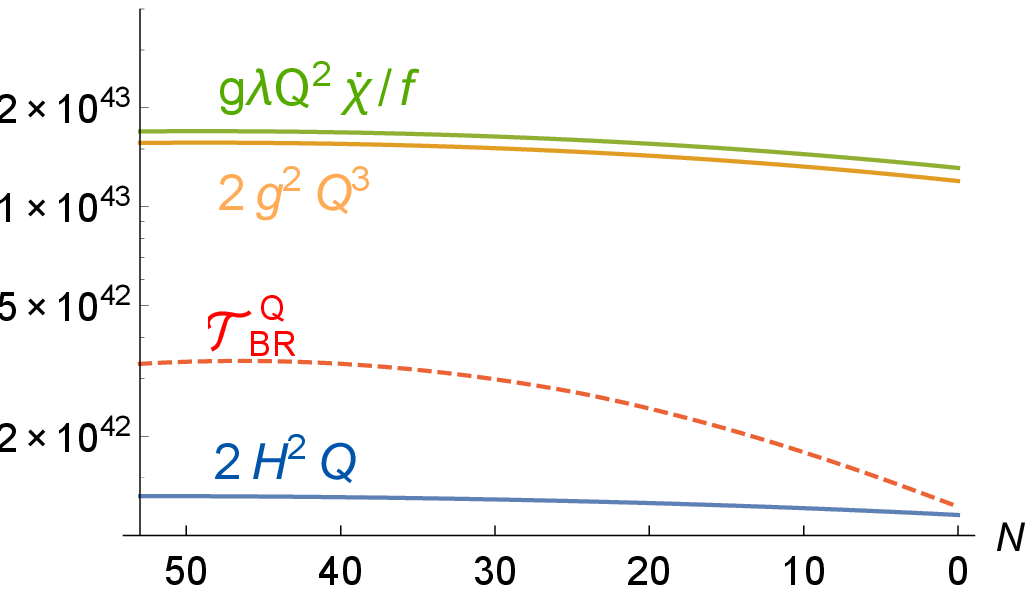}
  \hspace{5mm}
  \includegraphics[scale=0.65]{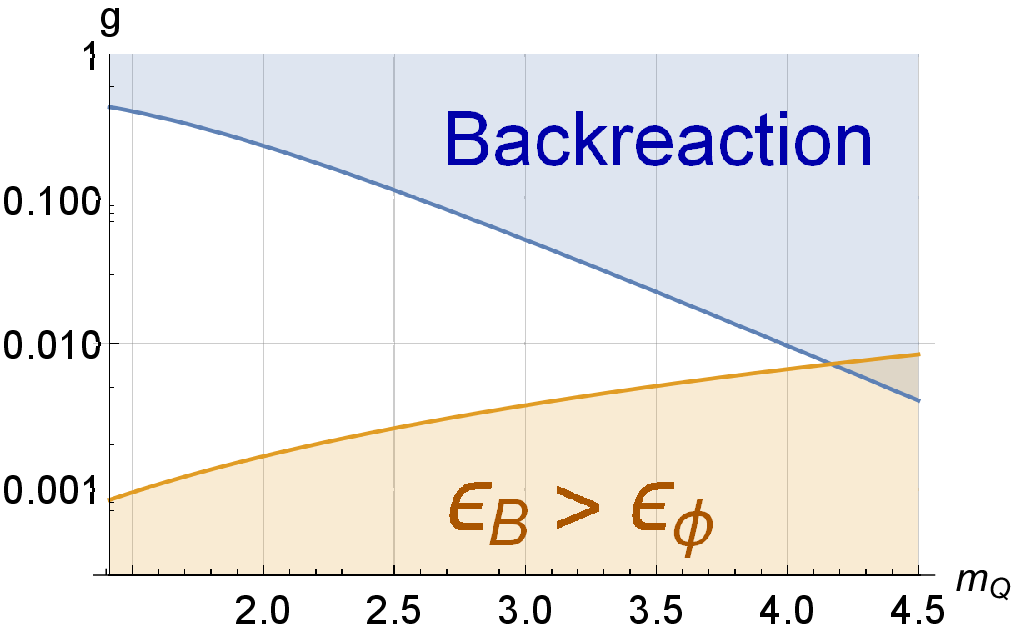}
  \caption
 {{\bf(Left panel)} The three leading terms in the EoM for $Q$, $g\lambda Q^2 \dot{\chi}/f$ (green), $2g^2 Q^3$ (yellow) and $2H^2 Q$ (blue) and the backreaction term from the tensor fluctuation $\mathcal{T}_{BR}^Q$ (red dashed). 
{\bf(Right panel)} The constraint on $g$ from the backreaction on the EoM for $Q$ (blue shaded region) and the assumption of single-field slow-roll inflation,
$\epsilon_B < \epsilon_\phi$ (yellow region). To study the parameters in those shaded region, one has to extend the treatment in this paper.}
 \label{backreaction}
\end{figure}
%

Before closing this section, we discuss the conditions required for the consistency of our treatment. The setup of our perturbation theory relies on there being only a small backreaction from the perturbations $t_R$ on the background equations and the requirement that  $\epsilon_B, \epsilon_{\chi}\ll \epsilon_\phi$.
We will label as a viable parameter space the regions where these conditions are met. However, we note here that this is a conservative approach as a treatment fully accounting for a stronger backreaction and
relaxing the slow-roll parameters hierarchy might well reveal additional acceptable domains in parameter space.\\
As we have seen, the amplification of the tensor mode fluctuation $t_R$ due to the (controlled) instability generates intriguing imprints in this model, but it does not occur at no cost.
The energy used to amplify $t_R$ is necessarily transferred from the background fields, $Q$ and $\chi$. In turn, $t_R$ can  backreact on the equations of motion for $Q$ and $\chi$.
Upon accounting for this effect, the following contributions appears as a correction to Eq.~\eqref{gauge-bck},
\begin{align}
\mathcal{T}_{BR}^Q\equiv\frac{g\xi}{3a^2} H  \int \frac{\dd^3 k}{(2\pi)^3} |t_R|^2
+ \frac{g}{3a^2} \int \frac{\dd^3 k}{(2\pi)^3} \frac{k}{a} |t_R|^2 \simeq \frac{gH^{3}}{12\pi^2} \left(\xi\mathcal{B}(m_Q)+
\tilde{\mathcal{B}}(m_Q)\right),
\label{def TQ}
\end{align}
with
\begin{equation}
\mathcal{B}(m_Q) = \int^{x_{\max}}_0 \dd x\, x \left|i^{\beta}\,W_{\beta,\alpha}(-2ix)\right|^2,\quad
\tilde{\mathcal{B}}(m_Q) = \int^{x_{\max}}_0 \dd x\, x^2 \left|i^{\beta}\,W_{\beta,\alpha}(-2ix)\right|^2.
\end{equation}
In the derivation we have used the analytic solution for $t_R$, Eq.~\eqref{analytic tR},
and introduced the UV cutoff $x_{\rm max}\equiv m_Q+\xi+\sqrt{m_Q^2+\xi^{2}}$ so as to encompass the main contribution in proximity of the horizon region. In the left panel of Fig.~(\ref{backreaction}), we plot the three leading terms in Eq.~\eqref{gauge-bck} as well as $\mathcal{T}_{BR}^Q$. In deriving the latter, the numerically obtained $m_Q$ is plugged into Eq.~\eqref{def TQ}.
One can see that $\mathcal{T}_{BR}^Q$ is indeed sub-leading in the case of the parameters chosen in Eq.~\eqref{parameter set 1}.
For one to neglect backreaction effects, $\mathcal{T}_{BR}^Q$ should be much smaller than the largest term in Eq.~\eqref{gauge-bck}, namely $g\lambda Q^2 \dot{\chi}/f$. This condition is rewritten as
\begin{equation}
\mathcal{T}_{BR}^Q \ll g\lambda \frac{Q^2 \dot{\chi}}{f}
\qquad\Longleftrightarrow\qquad
g\ll \left(\frac{24\pi^2 m_Q^2}{\mathcal{B}+\tilde{\mathcal{B}}/\xi}\right)^{1/2}.
\label{inequality1}
\end{equation}
 Upon using the relation
 $\xi\simeq m_Q+m_Q^{-1}$, which is valid in the slow-roll approximation, the r.h.s. can be evaluated as a function of $m_Q$; this is essentially how the no-go backreaction region is obtained in e.g. Fig.~(\ref{F and epB}),(\ref{Low H}). From a similar check performed on the equation of motion for $\chi$, one  derives a slightly weaker condition.\\
The tensor polarization $t_R$ also contributes to the Friedman equation, Eq.~\eqref{our Friedmann}.
The energy density of $t_R$ is given by
\begin{equation}
\rho_{t_R} = \frac{1}{a^4} \int \frac{\dd^3 k}{(2\pi)^3}
\left[\, \frac{1}{2} |\partial_\tau t_k^R|^2 + \left(\frac{k}{2} - \frac{m_Q}{\tau}\right)k |t_k^R|^2 \right]\simeq \frac{H^4}{8\pi^2} \mathcal{I}(m_Q).
\label{rho tR}
\end{equation}
with
\begin{equation}
\mathcal{I}(m_Q) =
\int_0^{x_{\max}} \dd x\, x^3 
\left[ \left| i^{\beta} \partial_xW_{\beta,\alpha}(-2ix)\right|^2+
\left(1+\frac{2m_Q}{x}\right) \left| i^{\beta} W_{\beta,\alpha}(-2ix)\right|^2\right].
\label{I def}
\end{equation}
Using this expression we have plotted $\rho_{t_R}$
in Fig.~(\ref{chi and rho}) as the purple dashed line. One can see that
$\rho_{t_R}$ gives only a sub-leading contribution.

Finally we discuss the condition, $\epsilon_B\ll \epsilon_\phi$.
In eq.~\eqref{EoM for epphi}, we have assumed that the main contribution to the slow-roll parameter $\epsilon_H$ comes from the inflaton sector, in the form of $\epsilon_{\phi}$. The dynamics may(not) be altered if this is not the case. 
Since $\epsilon_B>\epsilon_E, \epsilon_\chi$ in our case (see Fig.~\ref{epsilon and mQ}), it is sufficient to impose $\epsilon_B\ll \epsilon_\phi$.
We note in passing that this condition can actually be relaxed by modifying  Eq.~\eqref{EoM for epphi} to account for a larger  $\epsilon_B$. However, we stick here to dynamics for which this condition is to be met and rewrite it as
\begin{equation}
\frac{\epsilon_B}{\epsilon_\phi} =\frac{g^2}{8\pi^2 \mcP_\zeta m_Q^{4}}\ll1
\qquad\Longleftrightarrow\qquad
g\ll \sqrt{8\pi \mcP_\zeta}m_Q^2,
\label{inequality2}
\end{equation}
where we have used $\mcP_\zeta = H^2/(8\pi^2 \epsilon_\phi \Mpl^2)$.
In the right panel of Fig.~(\ref{backreaction}), we plot Eqs.~\eqref{inequality1}
and \eqref{inequality2}. In the right panel of Fig.~(\ref{F and epB}), the constraint of Eq.~\eqref{inequality1} is shown as the red shaded region.
For $g\lesssim 7\times 10^{-3}$, Eq.~\eqref{inequality2} gives a stronger restriction instead.

\section{Discussion}
\label{sec:discussion}

In \textit{Section} \ref{sec:our} we have provided a detailed study of the rich phenomenology of this model for some specific regions of the parameter space. It is essential that we gauge the extension of this viable and possibly testable parameters set. We will proceed in two different direction. First, we scan decreasing values of the Hubble parameter towards a lower scale inflationary regime. Secondly, we scale down the value of the coupling constant, $\lambda$, regulating the axion-gauge field interaction strength.

In typical single-field slow-roll models, a smaller Hubble scale comes with a smaller slow-roll parameter $\epsilon$ favoring an undetectable value for the tensor-to-scalar ratio. This is not the case for the theory under study. Indeed, the main source of gravitational waves generation relies here on the presence of SU(2) gauge fields.  On the scalar side too, a small value for $\epsilon$ in not in tension with the observed spectral index as the main contribution to latter is now due to the parameter $\eta$.  
%
\begin{figure}[tbp]
    \hspace{-2mm}
  \includegraphics[width=70mm]{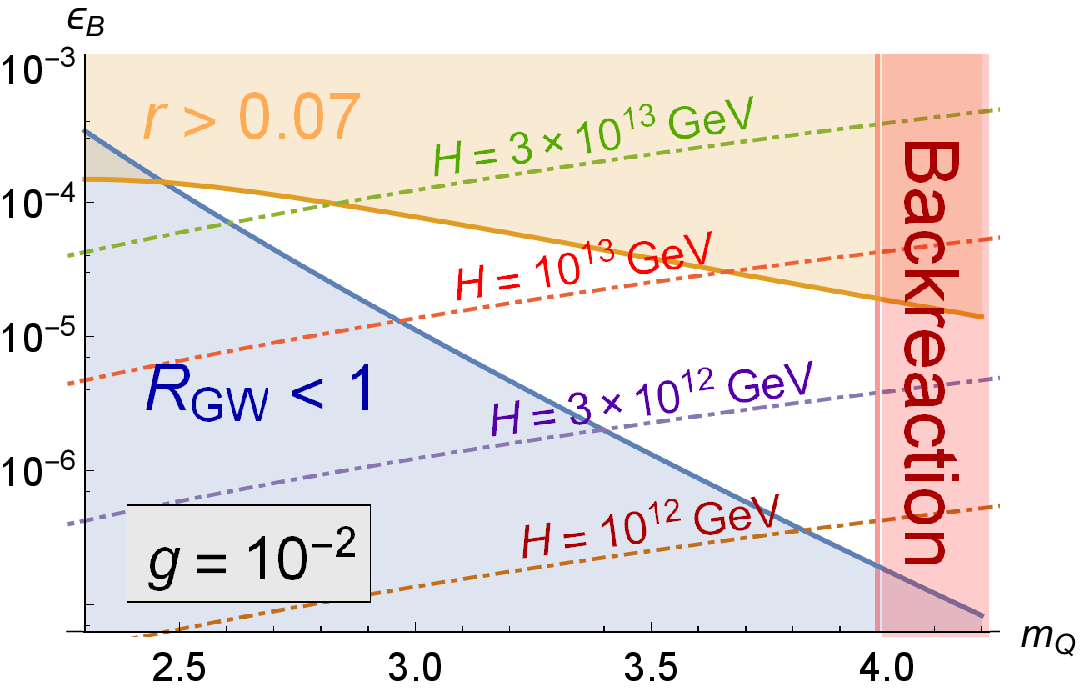}
  \hspace{5mm}
  \includegraphics[width=70mm]{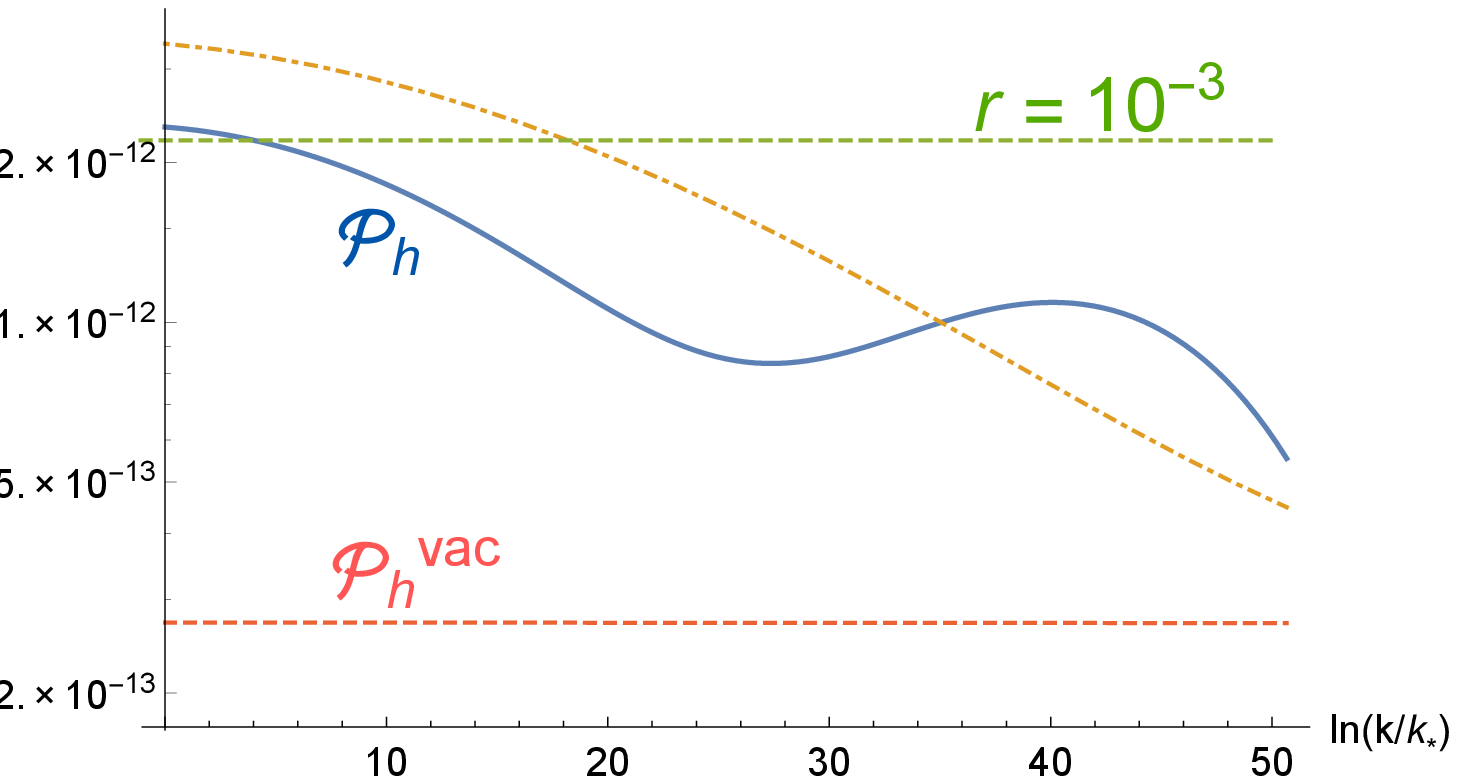}
  \caption
 {{\bf(Left panel)} Here the contour lines signalling $H=3\times 10^{13}\GeV$ (green), $10^{13}\GeV$ (red), $3\times10^{12}\GeV$ (purple), and $10^{12}\GeV$ (brown) are drawn on a plot reproducing the right panel of fig.~\ref{F and epB}. Due to the backreaction constraint, the lowest allowed $H$ with $\mcR_{GW}>1$ in our treatment is around $10^{12}\GeV$. 
{\bf(Right panel)} The GW power spectrum in the case with a low Hubble scale, $H_* = 2.82\times10^{12}\GeV$. While the vacuum GW is much smaller than the detectable level ($r=10^{-3}$), the sourced GW exceeds it around $k=k_*$. The other parameters are set as  $g=10^{-2}, \lambda=500,  \chi_*=5f/3=1.67\times10^{16}\GeV, \mu = 4.87\times10^{14}\GeV$.}
 \label{Low H}
\end{figure}
%

The most relevant effect of lowering $H$ on our dynamics is an increased initial value for   $Q$ and especially for $m_Q,\epsilon_{\chi}, \epsilon_B$. Through the background equations of motion this turns into a sharper, steeper time dependence of $m_Q$ resulting, if only $H$ is being varied, into a shorter-lived but more powerful chiral GW signal. It turns out the effects of a lowering Hubble scale can be undone by correspondingly decreasing the parameter $\mu$ in the axion potential. Intriguingly, this $H\leftrightarrow \mu$ balancing mechanism can in principle go on for a wide range of scales, with a parameter space that still supports a tensor-to-scalar ratio of the order $10^{-3}$ at CMB scales.

Instead, what, in our treatment, sets the lower bound on the Hubble scale, is the necessity to steer clear from regions in the parameter space where backreaction effects become too large and/or the perturbative description breaks down (see \cite{Peloso:2016gqs} for an in depth study of these matters in a somewhat similar setup). We have addressed the most pressing of such questions in \textit{Section} \ref{sec:backreaction} above\footnote{However, a full treatment of the bounds set by these requirements goes beyond the scope of this paper and will be the subject of a future work \cite{future1}}. In particular, this provides us with a lower bound on $H$ stemming from the combined requirements that (i) the background energy density be larger than the energy density due to the tensor mode perturbation $t_R$ and that (ii) the full eom for $Q, \chi$ is, again, dominated by terms which are background contributions and therefore not proportional to $t_{R}^2$. For the parameters set corresponding to the left panel Fig.(\ref{Low H}), one finds  $H\gtrsim 10^{12}\text{GeV}$ and the bound stays within the same order of magnitude as one navigates the viable parameter region.

\begin{figure}[tbp]
    \hspace{-2mm}
  \includegraphics[width=70mm]{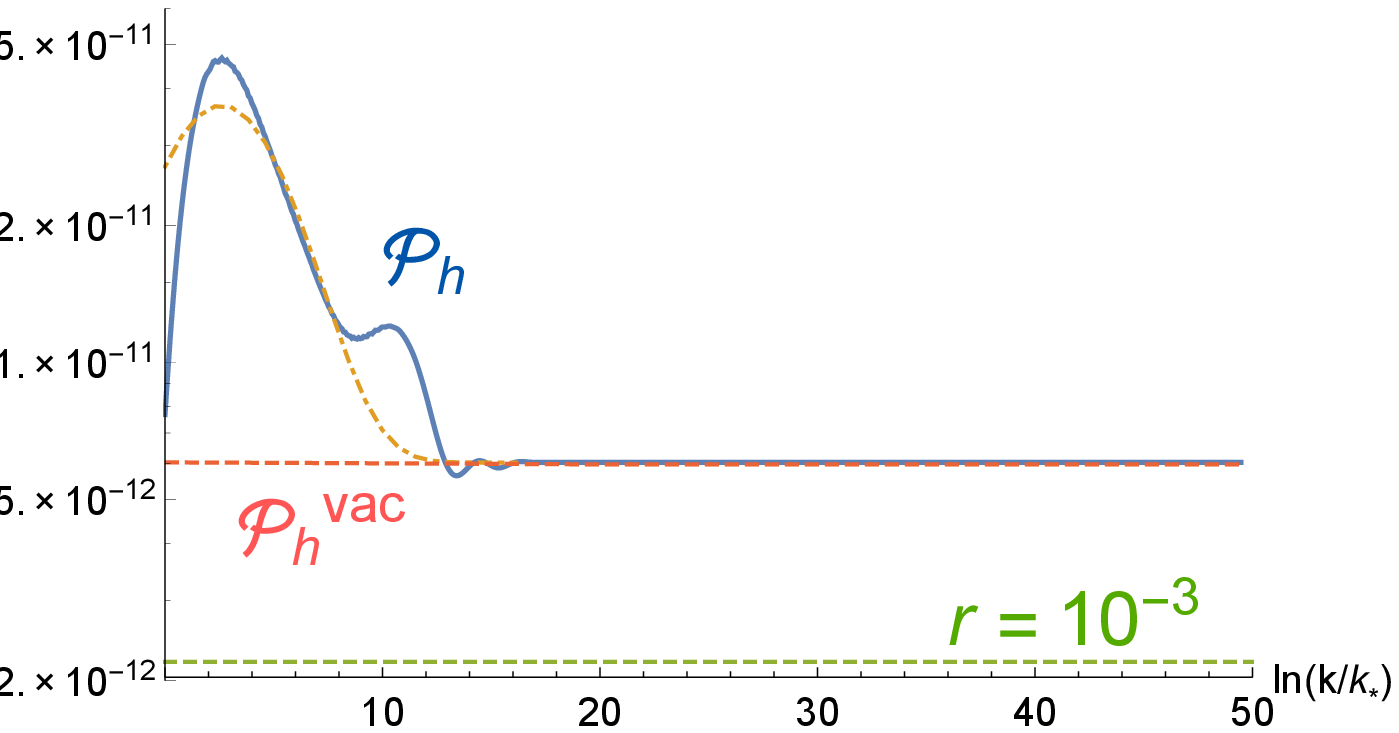}
  \hspace{5mm}
  \includegraphics[width=70mm]{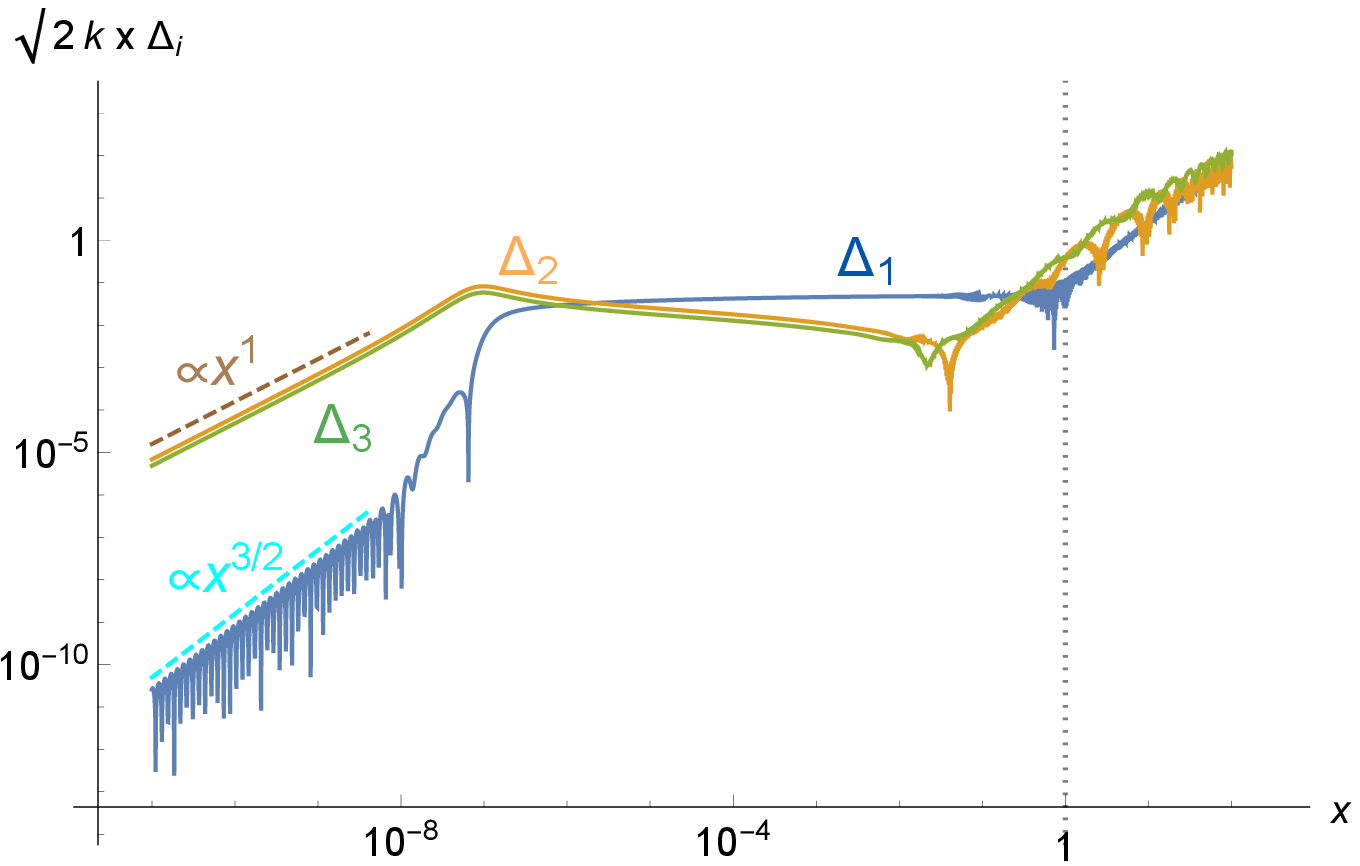}
  \caption
 {{\bf(Left panel)} The GW power spectrum in the smaller coupling constant $\lambda$ case. GW is enhanced only on the larger scale $\ln(k/k_*)\lesssim 12,$ because the slow-roll regime of $\chi$ lasts for a shorter time interval and $m_Q=\mathcal{O}(1) $ only for $60\lesssim N \lesssim 35$.  
{\bf(Right panel)} The time evolution of the scalar perturbations $\Delta_i (i=1,2,3)$. As the background fields decreases $\chi\propto a^{-3/2}$ and $Q\propto a^{-1}$, the corresponding perturbations also decay in the same way, $\Delta_1\propto x^{3/2}$ and $\Delta_{2,3}\propto x$. In this case, the curvature perturbation generated by  density perturbation, $\delta\rho_\chi$ and $\delta\rho_Q$, becomes negligible.
In both panels, we set $g=10^{-2}, \lambda=50, \chi_*=5f/4=6.32\times10^{15}\GeV,H_* = 1.32\times10^{13}\GeV, \mu = 1.1\times10^{15}\GeV$.}
 \label{Low lambda}
\end{figure}

We now move on to the allowed range for the parameter $\lambda$ and focus on whether a weaker coupling than necessary for CNI still supports an interesting phenomenology in our case. In the chromo-natural mechanism the importance of a sufficiently large  $\lambda$ is manifold. Most importantly, the duration of the inflationary era itself relies on the effective axion potential ability to accommodate a slow-roll phase. For the latter to last long enough and avoid hitting the Planckian regime  $f\sim M_{P}$,  one needs the scalar gauge degrees of freedom to efficiently slow down the axion field via the  $\lambda$-regulated coupling. Even more relevant to our setup is the fact that it is this very interaction that feeds and prevents the scalar gauge degrees of freedom $Q, U$ (and, indirectly, the tensors $t^{R/L}_{ij}$ as well) from decaying too early. Indeed, although we no longer need the axion to sustain inflation, the characteristic ``chiral" imprint on the tensor power spectrum, its magnitude and duration, is still based upon the existence of a sufficiently strong axion-gauge coupling. The difference is that now we have effectively decoupled these chiral signatures from the bulk of the inflationary dynamics\footnote{This includes the scalar spectral index whose leading contribution is now due to the inflaton field $\phi$.}. The axion no longer needs to slow-roll for e.g. sixty e-folds, a few are enough to generate, for example, interesting CMB imprints. As a consequence, the parameter space of the theory now admits smaller values for $\lambda$, from $500$ down to an order of magnitude of $\lambda \gtrsim 50$. Within the same $\lambda$ interval, the CMB scales tensor signal can support  both a red or a blue tilt and, as illustrated in Fig.~(\ref{Low lambda}), a tensor-to-scalar ratio as high as $10^{-2}$. 
%
%

\section{Conclusions}
\label{sec:conclusions}
In studying the inflationary era the standard lore is that the analysis of the scalar sector, in particular non--Gaussianities, is the most direct probe of inflationary dynamics. The tensor signal, if detected, typically provides a measure of the energy scale of inflation. Our setup lies somewhere outside this realm. Although the acceleration is driven by the inflaton, the sector most sensitive to additional field content is here the tensor one and the Hubble scale can no longer be directly inferred from the knowledge of the tensor-to-scalar ratio. Indeed, the mechanism generating the leading contribution to gravitational waves is not that of vacuum fluctuations. Rather, it relies on the presence of an axion-gauge fields coupling feeding the tensor gauge degrees of freedom and, ultimately, the tensor power spectrum. This being an alternative source of gravitational radiation, it is intriguing that it is also automatically endowed with a distinct signature such as chirality and a tensor spectral index spanning both red and blue values.

Both the vector and scalar degrees of freedom at play can be traced back to those of chromo-natural inflation. On the other hand, CNI is ruled out because, upon requiring from the axion a scalar spectral index compatible with the measured one, the model is forced to acquire too large a value for the tensor-to-scalar ratio. The tension with data is relaxed if an additional  field is introduced and put in charge of inflating. What was an inflating axion in CNI is now a slowly rolling field generating a distinct signature on e.g. CMB scales. Having abdicated the role of inflaton, the axion slow-roll phase can be much shorter whilst still delivering interesting imprints. As a consequence, the parameters space corresponding to those same degrees 
of freedom of CNI is now much wider, and spans, for example, a smaller value for the coupling constant $\lambda$.

In putting this model forward, we have been deliberately agnostic about the inflaton self-interactions. We required that $\epsilon \ll \eta \propto(n_s-1)$ and focussed on the ensuing phenomenology. This leaves ample room for a future embedding of the inflaton sector within a supergravity context. It would be of particular interest to probe the role of the inflaton field as a dilaton kinetically coupled to the gauge fields as the counterpart of the axion-gauge coupling \cite{recentCNI}d. One would ideally identify and study the K{\"a}hler potential at the origin of such a setup. More in general, the leading contribution to scalar non-Gaussianities will depend upon the choice of the specific inflationary potential.

As to post-inflationary dynamics, one should first recall that this model can accommodate for an axion that decays (together with the gauge d.o.f.s) during inflation. Notably, this corresponds to a viable parameter space domain that generates a detectable chiral gravitational waves signal. It also implies a reheating stage essentially equivalent to that following a single-field inflationary era. On the other hand, an exhaustive treatment calls for the investigation of all the inflaton and axion decay channels. We leave this, as well as the study of non-Gaussian signatures, to future work.  

Finally, we stress again that the viable parameter space for the theory has been obtained by placing the initial dynamics close to the CNI configuration and by avoiding the domains supporting large backreactions effects. Both choices lead to a simplification of the analysis but do not strictly rule out the portions of the parameters  excluded from our reach. It would indeed be very interesting to relax these assumptions pursue a more adventurous  route which may reveal an even richer dynamics.


 \section*{Acknowledgements} 
  We are grateful to Eiichiro Komatsu for very fruitful discussions. E.D. acknowledges support by the DOE under grant No. de-sc0008016. M.F.  is supported in part by NSF grant PHY-1068380. The work of TF is partially supported by the JSPS Postdoctoral Fellowships for Research Abroad, Grant No. 27-154. ED would like to thank SITP and Stanford University for warm hospitality whilst this work was being completed.


\end{document}